# OrbGNN: A Wave function-based Machine Learning Interelectronic Representation

Brody Quebedeaux,[1] Shahzad Akram,[1] Markus Reiher,[2] Konstantinos D. Vogiatzis[1,*]

[1]*Department of Chemistry, University of Tennessee, Knoxville, Tennessee 37996, United States*

[2]*Department of Chemistry and Applied Biosciences, ETH Zurich, Vladimir-Prelog-Weg 2, Zurich 8093, Switzerland*

[*]Corresponding Author: kvogiatz@utk.edu

## Abstract

Machine learning interatomic potentials (MLIPs) have become emerging tools in molecular modeling and computational chemistry. By learning high-dimensional potential energy surfaces from quantum chemical data, MLIPs enable accurate and efficient predictions of structural, thermodynamic, and dynamical properties. However, such models have limitations in predictions of electronic properties and the effects of static electron correlation due to their lack of electronic structure information. This work presents OrbGNN, an electronic structure graph architecture analogous to molecular graph and MLIP frameworks, where pair-orbital interactions constitute the graph representation, while orbital entanglement encodes the connectivity between them. By embedding information derived from orbital correlation metrics directly into the graph topology, OrbGNN provides a compact representation of a molecule's orbital landscape and electron correlation patterns. Analysis of the feature space of the orbital graph demonstrates the robustness of the model. The model is evaluated for the dissociation of nitrogen and for a larger dataset of diatomic molecules. Finally, the OrbGNN model is applied to a set of octahedral iron(II) complexes to predict spin-state energy gaps.

# Introduction

Machine learning interatomic potentials (MLIPs)[1] are reshaping the landscape of molecular modeling and computational chemistry.[2-4] At the heart of computational chemistry lies a fundamental conflict: the trade-off between speed and accuracy. MLIPs have emerged to bridge this gap, engaging data-driven techniques to provide first-principles accuracy to classical molecular simulations through predictions of highly accurate forces. In addition to the major advances in accurate simulations, MLIPs provide a framework for neural network model architectures. Modern advances in MLIPs have highlighted their ability to map the atomic and molecular structure of a system to its potential energy surface (PES). Modern MLIP architectures, when combined with large and diverse chemical datasets, enable for the development of universal MLIPs, aiming to extrapolate their predictive capabilities across a large number of chemical systems.

Viewing the current landscape of MLIPs, there are methods ranging from surrogate models aimed at specific systems and properties, to foundational models made to tackle a large variety of chemical systems.[1, 5-13] Still, the lack of explicit electronic-structure information, limited extrapolation, and inadequate descriptions of long-range physics.[2-4] Efforts to improve predictions on these frontiers are ongoing, incorporating long-range interactions into existing MLIPs[14-16] and models focusing on electronic structure.[17] For electronic structure, in particular, current efforts focus on learning wavefunctions from data-driven schemes,[18-21] creating neural network wavefunctions,[22-25] and achieving higher fidelity accuracy, such as at the coupled-cluster (CCSD) level, from lower fidelity methods, such as Hartree-Fock (HF) or density functional theory (DFT).[26] While these efforts reach broadly across wavefunction-based methods, a subset of this field focuses on multi-reference methodologies for static electron correlation (also referred to as strong electron correlation). Static correlation denotes electron correlation that arises when multiple electronic configurations are needed to describe a molecule's electronic structure, and arises in transition metal complexes, actinides, bond dissociation processes, and excited states, where the multiconfigurational character of the wavefunction cannot be captured by single-reference approaches. Full configuration interaction[27] (FCI) provides the exact multi-configurational energy for a system within a given orbital basis, but it is massively limited in its applicability with respect to system size. Multi-configurational self-consistent field (MCSCF) theory is a branch of electronic structure theory that delivers methods to approximate FCI while being capable of calculating properties for much larger systems. In this realm, a standard ansatz is complete active space self-consistent field (CASSCF)[28] which introduces active spaces, where a FCI calculation is carried out over a reduced set of strongly correlated orbitals, the so-called active orbitals. While CASSCF and its extensions such as RASSCF[29] and GASSCF[30] provide systematically improvable descriptions of static correlation, their computational cost scales steeply with active space size, rendering high-throughput application or large-scale screening of chemical space impractical. This bottleneck is particularly acute for transition metal chemistry,[31] where the interplay between metal d-orbital manifolds, ligand field effects, and spin state energetics demands both large active spaces and accurate treatment of correlation to achieve quantitative results. Alternatives to these methods include selective CI methods, where configurations are chosen on the fly in order to truncate the FCI wavefunction,[32-35] full configuration interaction quantum Monte Carlo (FCIQMC),[36] adaptive shift FCIQMC,[37] and the density matrix renormalization group (DMRG) algorithm.[38, 39] DMRG optimizes matrix product states (MPS) in a sweeping algorithm to overcome system size limitations in traditional MCSCF methods. In recent years, the development of ML architectures for advancing multireference methodologies has emerged as an active research direction. Existing ML models have aimed to bridge the gap between accuracy and efficiency in wavefunction based methods.[26, 40-43] These wavefunction based ML algorithms have begun contributing to the development of efficient MLIP representations.[17, 21, 44]

In this work, we describe a wavefunction based model, OrbGNN, inspired by the architecture of modern MLIPs. Analogous to the formation of a molecular graph in graph neural network (GNN) based MLIPs, we form orbital graphs. These graphs encode molecular orbitals as nodes and employ wavefunction-derived features to craft the edges between them, while features describing orbital connectivity quantify the connectivity between orbital pairs. Orbital connectivity measures the extent to which molecular orbitals are quantum mechanically coupled through electron correlation, revealing how their occupation probabilities become interdependent beyond a single-determinant description. As such, it provides a rigorous, quantitative assessment of multireference character by capturing the correlated fluctuations of orbital occupations that are absent at the mean-field level. Unlike energy-based diagnostics, connectivity metrics probe the internal structure of the many-electron wavefunction directly, offering a physically transparent measure of correlation strength and its distribution across orbital subspaces. Here, we employ orbital connectivity as an intrinsic, wavefunction-derived descriptor that encodes the quantum interdependence of orbitals and thus serves as a compact yet information-rich representation of electronic structure complexity. Three different metrics of orbital connectivity have been implemented in OrbGNN, namely mutual information,[45] mutual correlation,[46] and the two-electron cumulant.[47] These metrics were chosen due to their distinct differences and coverage of the ways in which connectivity is quantified. While mutual information is a direct measure of orbital-pair entanglement in a quantum information theory context, the two-electron cumulant provides a detailed look at two-body interactions directly from the two-electron reduced density matrices, and mutual correlation provides local orbital pair information. With the connectivity and a variety of other electronic structure properties described in this paper, OrbGNN shows chemical accuracy when predicting energies for a variety of diatomic systems and it predicts spin gaps of Fe (II) octahedral complexes to ~1 kcal mol$^{-1}$ error.

# Methods

## Orbital Correlation Metrics

**Mutual Information:** Mutual information,[45, 48-50] a property derived from quantum information theory, describes the interaction between pairs of orbitals through the von Neumann entropy $s$. The one-orbital von Neumann entropy, a function of the eigenvalues of the one-orbital reduced density matrix (oo-RDM), is calculated as

$$s(1)_i = -\sum_{\alpha=1}^{4} \omega_{\alpha,i} \ln \omega_{\alpha,i} \quad (1)$$

where $\omega_{a,i}$ are the eigenvalues of the oo-RDM and $\alpha$ refers to the occupations possible for a spatial orbital, 0, 1 ($\alpha$ or $\beta$ spin), or 2 electrons. This can be interpreted as the entanglement of orbital $i$ with the rest of the environment.

Using the two-orbital reduced density matrix $\omega_{\alpha,i,j}$ (to-RDM), the two orbital von Neumann entropy is calculated as

$$s(2)_{ij} = -\sum_{\alpha=1}^{16} \omega_{\alpha,i,j} \ln \omega_{\alpha,i,j} \tag{2}$$

where the index $\alpha$ now runs over the possible occupations of two orbitals. Finally, the two-orbital mutual information, denoted as $I_{ij}$, quantifies the correlation between orbitals *i* and *j* by removing their individual entropy contributions from the two-orbital entropy.

$$I_{ij} = \frac{1}{2}\left[s(1)_i + s(1)_j - s(2)_{ij}\right]\left(1 - \delta_{ij}\right) \tag{3}$$

If there is no entanglement between orbitals *i* and *j*, the two-orbital von Neumann entropy is simply the sum of their individual one-orbital von Neumann entropies

$$s(2)_{ij} = s(1)_i + s(1)_j \tag{4}$$

therefore, the mutual information is taken as the reduction of $s(2)_{ij}$ by the sum of $s(1)_i$ and $s(1)_j$. Although one- and two-orbital entropies – and hence, mutual information – are readily available in the framework of the density matrix renormalization group (DMRG) algorithm, they can also be calculated from the standard one- and two-body reduced density matrices.[51, 52]

**Cumulant:** The two-cumulant[47] is derived first from the one- and two-body reduced density matrices (1-RDM and 2-RDM, respectively):

$$\gamma_r^p = \langle\psi|a_p^\dagger a_r|\psi\rangle \tag{5}$$

and

$$\gamma_{rs}^{pq} = \langle\psi|a_p^\dagger a_q^\dagger a_s a_r|\psi\rangle \tag{6}$$

where $a_p^\dagger$ and $a_p$ are creation and annihilation operators, respectively. To calculate the two-cumulant, the 2-RDM isolates correlation that do not factorize into products of 1-RDMs

$$\lambda_{rs}^{pq} = \gamma_{rs}^{pq} - \left(\gamma_r^p \gamma_s^q - \gamma_s^p \gamma_r^q\right) \tag{7}$$

thereby deriving the true two-particle correlations from the 2-RDM. In this work, the two-cumulant is used to quantify correlations between orbital pairs, taken as

$$\lambda_{ij} = \sum_{rs} {\lambda_{rs}^{ij}}^2 \tag{8}$$

where we quantify the cumulant-based correlation with respect to the environment around orbital pairs.

**Mutual Correlation:** Mutual correlation, introduced by Evangelista,[46] is a quantification of subsystem correlations through partitioning of the cumulant norm. For the purposes of this work, we employ the pair mutual correlation for orbital pairs:

$$\mathcal{M}_{PQ} = (\Lambda^2)_{QQ}^{PQ} + \frac{1}{2}(\Lambda^2)_{QQ}^{PP} + (\Lambda^2)_{PQ}^{PQ} + (\Lambda^2)_{QP}^{PP} \tag{9}$$

where,

$$(\Lambda^2)_{RS}^{PQ} = \left|\lambda_{R\uparrow S\uparrow}^{P\uparrow Q\uparrow}\right|^2 + \left|\lambda_{R\uparrow S\downarrow}^{P\uparrow Q\downarrow}\right|^2 + \left|\lambda_{R\uparrow S\downarrow}^{Q\uparrow P\downarrow}\right|^2 + \left|\lambda_{R\uparrow S\downarrow}^{P\uparrow Q\downarrow}\right|^2 + \left|\lambda_{R\uparrow S\downarrow}^{Q\uparrow P\downarrow}\right|^2 + \left|\lambda_{R\downarrow S\downarrow}^{P\downarrow Q\downarrow}\right|^2 \tag{10}$$

The *P,Q,R,S* indices correspond to spatial molecular orbitals, and the up, down arrows to alpha and beta spin orbitals, respectively. This decomposition approach to the two-cumulant norm provides a way to discern how correlation effects are distributed across a system, as it is not a property measuring the sum of correlation effects. In an orbital sense, the mutual correlation gives insight into the correlation between specifically orbitals $p$ and $q$, without interactions with their environment.

In summary, mutual information captures the entanglement between orbitals as true particle correlations derived from one and two orbital entropies. Described initially in quantum information theory, one-orbital entropies provide the quantitative measurement of entanglement between an orbital and the full orbital environment. Two-orbital entropies provide the measure of entanglement between an orbital pair and its environment. The derivation of these orbital entropies is built from the calculation of orbital reduced density matrices (o-RDMs). Ultimately, mutual information provides the true correlations in the two-orbital entropies that do not reduce down to individual one-orbital entropies. The norm of the cumulant describes a metric of correlation between orbitals. Unlike the mutual information, the cumulant norm is not a true metric of orbital entanglement, as it is not based on entropies. However, the cumulant is a property derived from the 2-RDM, and when taken as a norm becomes size extensive. This allows the cumulant norm to be taken as a correlation measure. Mutual correlation is a partitioning of the cumulant norm, taken as a localized measure of correlation. By a clever partition, the mutual correlation can derive the specific correlation between orbital pairs. As a general overview, the cumulant norm is taken as an aggregate correlation weight between orbital pairs, and the mutual correlation is taken as a localized linear proxy of the pairwise entanglement described by the entropy-focused mutual information.

In this work, all orbital correlation metrics are generated by DMRG calculations (*vide infra*).

## OrbGNN Architecture

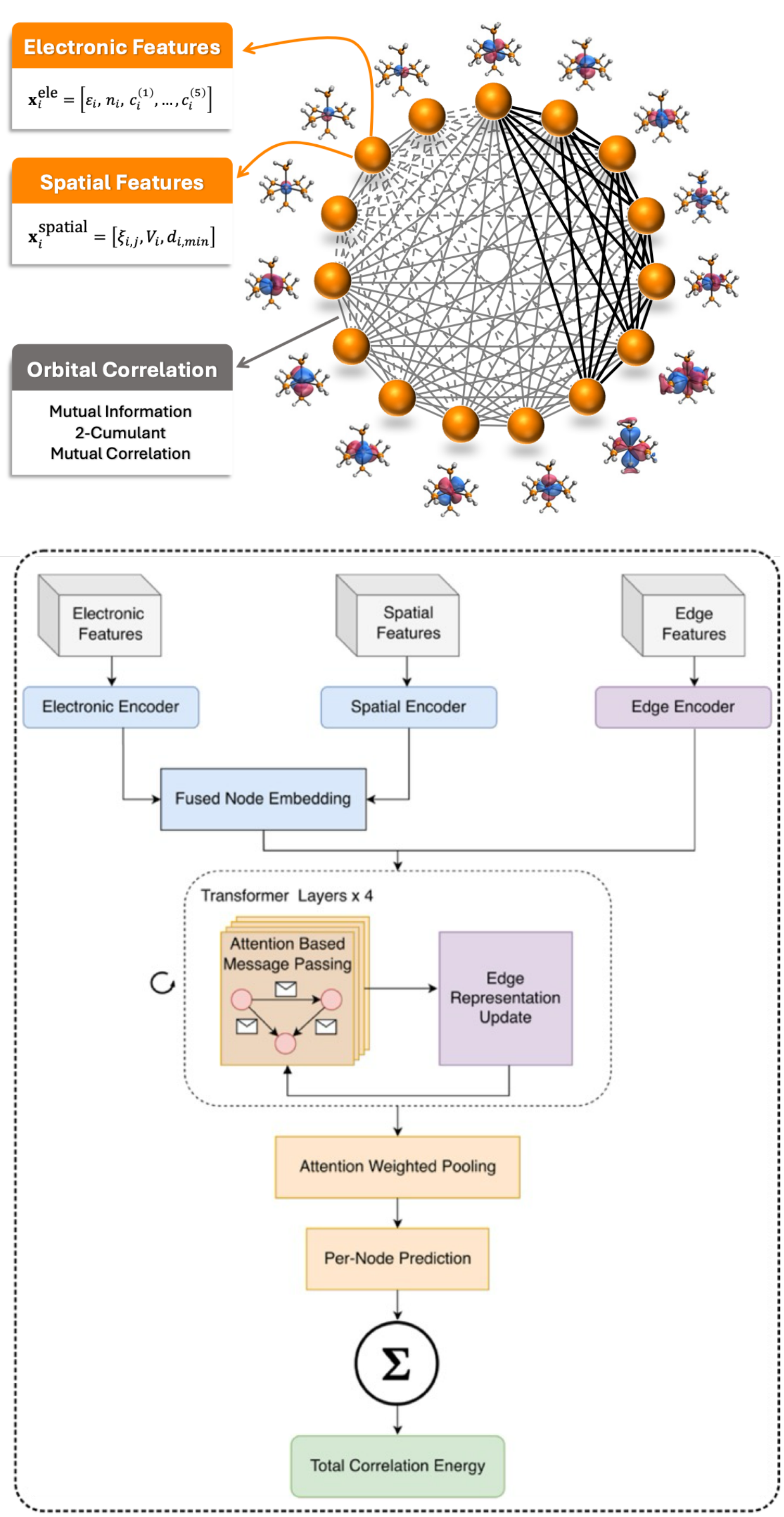

**Figure 1.** OrbGNN model overview. Top: Molecular orbital graph, where each node encodes electronic and spatial orbital information, and each edge encodes orbital correlation information. Bottom: model architecture with node-related operations shown in blue, edge-related operations in purple, and specific architectural components in orange.

OrbGNN is a message-passing graph neural network that operates over orbital graphs (Figure 1, top), in which nodes represent molecular orbitals and edges encode pairwise connectivity between them. Node initialization follows a two-stream design that separates electronic and spatial information before fusing them into a unified representation. The semantic stream processes normalized electronic features,

$$\mathbf{x}_i^{\text{ele}} = \left[\varepsilon_i, n_i, c_i^{(1)}, \dots, c_i^{(5)}\right] \tag{11}$$

comprising the Hartree–Fock orbital energy $\varepsilon_i$, the DMRG natural orbital occupation number $n_i$, and the five largest atomic orbital contributions $c_i^{(1)}, \dots, c_i^{(5)}$ to each molecular orbital, through a linear projection into a latent space of dimension $H$, where $H$ is the hidden embedding dimension hyperparameter. In parallel, the spatial stream encodes two physically motivated descriptors that are deliberately excluded from normalization to preserve their physical meaning

$$\mathbf{x}_i^{\text{spatial}} = \left[V_i, d_{i,min}\right] \tag{12}$$

where $d_{i,\text{min}}$ is the distance from the orbital centroid to the nearest nucleus, and $V_i$ a Gaussian-weighted nuclear charge potential felt by each orbital, summed over nuclei $A$ with charge $Z_A$:[43]

$$V_i = \sum_A Z_A \exp\left(-\frac{r_{iA}^2}{2\sigma^2}\right) \tag{13}$$

The centroid distance is expanded into a 16-dimensional Gaussian radial basis function (RBF) representation with a cosine cutoff envelope, allowing the model to resolve distinct distance regimes, such as metal first-shell versus ligand-centered orbitals, without requiring the network to learn this structure from a single scalar. The weighted nuclear charge is passed through a dedicated two-layer multilayer perceptron rather than being $z$-scored, preserving both the sign and absolute scale of the electrostatic environment, which carries chemically meaningful information that standardization would destroy. The outputs of the two streams are concatenated and projected through a fused linear layer with layer normalization to produce the initial node embedding of dimension $H$.

A schematic representation of the OrbGNN architecture is shown at the bottom of Figure 1. Edge features encoding the pairwise orbital interactions through orbital correlation metrics and the HF orbital energy difference are projected into the same hidden dimension through a linear encoder before message passing begins. Each message-passing layer employs a graph transformer architecture based on TransformerConv[53], in which multi-head attention over the graph topology,

$$\alpha_{ij} \propto \exp\left(\frac{\left(W_Q h_i\right)^{\top}\left(W_K h_j + W_E e_{ij}\right)}{\sqrt{d}}\right) \tag{14}$$

is combined with bidirectional gated message aggregation. Specifically, messages are computed separately from source to destination and from destination to source, each modulated by a learned sigmoid gate,

$$g_{j \to i} = \sigma\left(W_g; [; h_i \ \ h_j \ \ e_{ij}]\right) \tag{15}$$

that controls the contribution of each message to the receiving node. This bidirectional scheme allows the model to distinguish the directed influence of orbital interactions rather than treating them as symmetric. Following aggregation, node representations are updated through a position-wise feed-forward network with SiLU[54] activations, and edge representations are simultaneously refined using a three-way concatenation of the updated source node, destination node, and current edge features. Residual connections from the initial node embedding are carried through each layer via a learned scalar gate $\beta_i$,

$$h_i^{(\ell)} \leftarrow \beta_i \, h_i^{(\ell)} + (1 - \beta_i) h_i^{(0)} \tag{16}$$

allowing the model to retain the original orbital identity information throughout the depth of the network. A spin state embedding is added to all node representations after the final message-passing layer, enabling the model to modulate its energy predictions based on the overall electronic configuration of the system. The total energy is obtained by an attention-weighted sum of per-node scalar readouts,

$$\hat{E} = \sum_{i \in V} a_i \ y_i, \ a_i = \mathrm{softmax}_{i \in V} \left( w_a^{\top} h_i^{\mathrm{final}} \right) \tag{17}$$

where the attention weights are computed from the final node representations and normalized within each graph. The loss is derived as the Huber loss between the calculated and predicted electronic energies.

$$\mathrm{L} = \frac{1}{\mathrm{N}} \sum_{\mathrm{n}=1}^{\mathrm{N}} \mathrm{z}^{(\mathrm{n})}, \mathrm{z}^{(\mathrm{n})} = \begin{cases} \dfrac{0.5\left(\hat{\mathrm{E}}^{(\mathrm{n})} - \mathrm{E}_{\mathrm{true}}^{(\mathrm{n})}\right)^2}{\beta} & \left|\hat{\mathrm{E}}^{(\mathrm{n})} - \mathrm{E}_{\mathrm{true}}^{(\mathrm{n})}\right| < \beta \\ \left|\hat{\mathrm{E}}^{(\mathrm{n})} - \mathrm{E}_{\mathrm{true}}^{(\mathrm{n})}\right| - 0.5\beta & \text{otherwise} \end{cases} \tag{18}$$

**Target Selection**

To generate the target properties for this study, multi-reference quantum chemical calculations were performed, with specific methodology tailored to each molecular system. For the diatomic systems, reference output targets were computed using the complete active space self-consistent field (CASSCF) method, as detailed in the Computational Details section. The target values were defined as the difference between the CASSCF and Hartree-Fock (HF) energies ($E_{\mathrm{CASSCF}} - E_{\mathrm{HF}}$), thereby isolating the energetic contribution of static electron correlation.

Conversely, for the more complex iron systems, the complete active space second-order perturbation theory (CASPT2) was employed. Here, the targets were established as the difference between the CASPT2 and HF energies ($E_{\mathrm{CASPT2}} - E_{\mathrm{HF}}$). Because CASPT2 introduces dynamic correlation to the multi-configurational CASSCF reference, this subtraction provides a comprehensive measure of the total electron correlation energy for each complex. This allows for a test of the model's capability to learn dynamic correlation through the network, without direct features to describe orbitals outside of the active space.

## Input Data Generation

DMRG calculations for data creation in the training of this model were performed with PySCF[55] and the Block2 code,[56] and the ANO-RCC-MB and ANO-RCC-DZ[57] basis sets. The orbital connectivity metrics were calculated using natural orbitals obtained from the DMRG-CI procedure, wherein the one-particle reduced density matrix is diagonalized from a fixed-orbital state. A modest bond dimension of 100 for the feature creation was selected for all calculations of diatomic molecules. The bond dimension was chosen to ensure that the input data required by the surrogate model can be generated at low computational cost. Benchmark calculations showed no significant differences in the performance of OrbGNN models trained using DMRG data obtained with smaller versus larger bond dimensions (see Supplementary Material). For all the calculations with diatomic molecules, the target correlation energies were computed at the CASSCF level with a full valence active space. The nomenclature CAS($n$,$m$) is used in analysis, where $n$ in the number of electrons and $m$ the number of correlated orbitals in the selected active space. For example, calculations for the $N_2$ molecule were performed with an active space of ten electrons in the eight valence molecular orbitals, or CASSCF(10,8).

The same DMRG settings were used to generate data for the Fe(II) complexes, with the exception of a slightly larger bond dimension of 200. Target values were computed at the CASPT2/cc-pVTZ[58] level of theory using the OpenMOLCAS[59] program package. The cc-pVTZ basis set was selected to maintain consistency with prior literature.[60] To construct the training dataset for the iron complexes, several CAS of increasing size were considered. Starting from a minimal CAS(6,5) comprising the five near-degenerate Fe 3d orbitals, the active space was systematically expanded to CAS(6,10) by including the second Fe d shell, CAS(10,7) by adding two ligand orbitals with antibonding character relative to the Fe center, and CAS(10,12) by including both the second Fe d shell and the antibonding ligand orbitals.

## Data Generation

For demonstrating the capabilities of OrbGNN, we have trained models using different datasets. First, we generated 500 $N_2$ geometries sampled across the potential energy surface over the bond length range $R = 0.7 – 2.8$ Å. A data split for training of 80/10/10 was chosen for train, validation, and testing data, respectively. We also explored different orbital-correlation metrics for defining the graph connectivity, including mutual information, the 2-cumulant, and mutual correlation. We then trained OrbGNN on a dataset of 18 different diatomic molecules (CF, CC, BeF, CN, BeN, $O_2$, BN, BC, $N_2$, LiN, $Be_2$, BF, $B_2$, BBe, BeC, BeO, BLi, BO), with 100 geometries generated for each molecule over the bond length range $R = 0.7 – 2.8$ Å, and with a data split of 80/10/10 for train, validation, and testing, respectively. Finally, training data for the octahedral iron complexes were generated by constructing 25 geometries in which the Fe-L bonds were systematically varied at evenly spaced intervals around their equilibrium bond lengths, as listed in the Supplementary Material. For the purposes of this study, five different octahedral Fe(II) complexes were considered with L = $NH_3$, CO, HCN, CNH, and $PH_3$. A data split of 90/10 for the training and testing data was chosen for this application.

## Results

### Comparison of Orbital Correlation Metrics

To demonstrate OrbGNN's ability to process orbital graphs and obtain correlation energy from a strongly correlated system, we first analyze model performance with respect to the dissociation of the dinitrogen molecule. As it is one of the main facets of this work, the analysis of the behavior of the orbital connectivity features was conducted by tracking the values of these features across the dissociation curve.

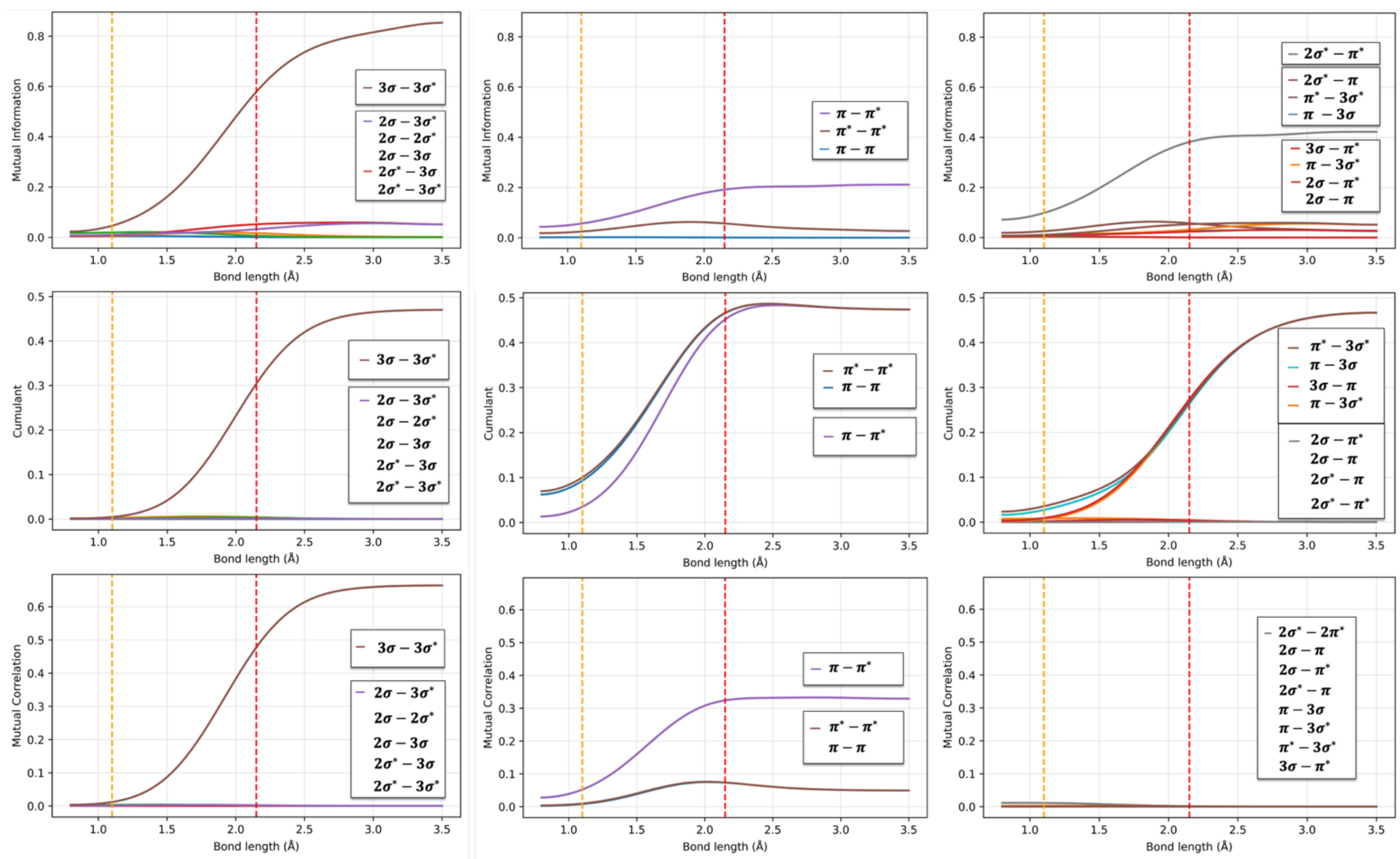

**Figure 2.** Behavior of orbital connectivity metrics across the dissociation of nitrogen. The graphs are divided into σ-σ interactions (left column), π-π interactions (central column), and mixed σ-π interactions (right column). Metrics computed from the mutual information, the two-cumulant, and the mutual correlation are shown in the top, middle, and bottom rows, respectively. The dashed red line in each graph corresponds to the $T_1$ diagnostic from coupled-cluster theory, a commonly used diagnostic of multireference character. The dashed yellow line indicates the equilibrium bond length.

Figure 2 demonstrates metric-dependent differences in the dissociation behavior of molecular nitrogen, where a general increase in connectivity is observed across all orbital pairs as the bond is elongated. This trend is consistent with the established electronic structure of nitrogen dissociation, wherein the $\pi$-manifold approaches degeneracy with increasing bond length, leading to enhanced static correlation. The two-cumulant and the mutual correlation exhibit closely aligned qualitative trends throughout the dissociation coordinate, although the two-cumulant consistently attains larger absolute magnitude. This behavior is consistent with their formal definitions; the two-cumulant tensor quantifies orbital-pair correlation relative to the full many-electron environment, retaining contributions that reflect coupling between the target pair and the remainder of the system, whereas the mutual correlation is partitioned to isolate intrinsic pairwise correlations by systematically excluding these environmental contributions. Dissociation profiles computed using the mutual information display qualitative trends similar to those obtained from the two-cumulant, albeit on a different numerical scale. This agreement arises from their shared sensitivity to correlations embedded within the complete orbital environment. However, quantitative discrepancies emerge due to their distinct mathematical origins: the two-cumulant is constructed from the two-electron RDM, whereas the mutual information is derived from o-RDMs, leading to systematic differences in magnitude along the dissociation coordinate.

Analyzing specific differences in both the qualitative and quantitative features of the plots yields distinct physical insights into the individual orbital spaces. Regarding the $\sigma$ interactions (Figure 2, left column), all metrics reach a consensus on their character, with the $3\sigma$-to-$3\sigma^*$ interaction dominating to reflect the electron excitation from the bonding to antibonding orbital necessary to cleave the nitrogen bond. In contrast, the metrics behave differently within the $\pi$ interactions (Figure 2, central column). The two-cumulant shows a convergence of all interaction types to the same value; because the cumulant metric is a summation across all four indices of the 2-RDM, this convergence indicates that as the molecule dissociates, the correlation in the $\pi$-manifold begins to be shared globally across all orbitals. Conversely, the mutual correlation provides a localized view of correlation between orbital pairs, allowing the three interactions to be resolved independently. The $\pi$-to-$\pi^*$ pair tracks the excitation of an electron required to activate the triple bond of the dinitrogen molecule. The $\pi$-to-$\pi$ and $\pi^*$-to-$\pi^*$ orbital correlation metrics provide smaller entanglement contribution, mostly comprised of small deviations in the orbital occupations. Mutual information tracks similar trends to mutual correlation but uniquely identifies the full dissociation of the nitrogen bond, marked by a decrease in entanglement around 1.9 Å. Because mutual information is based on orbital occupations derived from o-RDMs, this drop-off captures the onset of the dissociative regime of the molecule and the corresponding shift into two isolated nitrogen atoms.

Finally, the mixed-character interactions (Figure 2, right column) reveal structural variations depending on the chosen metric. For the mutual correlation, cross-orbital pairs, specifically those that do not share the same angular momentum quantum number L, exhibit a notable lack of localized correlation. This absence stems from electronic transition rules that disallow electron excitations between orbitals of the same parity. Meanwhile, the two-cumulant displays a convergence behavior analogous to its performance in the $\pi$ pairs. Most notably, the mutual information captures a maximum entanglement between the $2\sigma^*$-to-$\pi^*$ orbitals, an outcome that is not immediately intuitive from a classical chemical perspective. This elevated entanglement occurs because when an electron is excited from the $\pi$-to-$\pi^*$ orbital, the $2\sigma^*$ orbital experiences fluctuations in its electron occupation. Because mutual information is calculated directly from the o-RDMs, it sensitively detects these occupation fluctuations, resulting in the observed increase in entanglement.

**OrbGNN: Diatomic Molecules**

After analyzing the orbital connectivity metrics across the dissociation of the nitrogen molecule, we trained OrbGNN on data generated solely by dinitrogen molecules to recreate the $N_2$ dissociation curve using each of these metrics separately as edge features. Figure 3 (top) presents the OrbGNN prediction errors as a function of bond length for the three orbital graph descriptors (mutual information, two-cumulant norm, and mutual correlation). All three descriptors achieve mean absolute errors (MAEs) within sub-chemical accuracy (less than 1kJ mol$^{-1}$ or 0.38 m$E_h$), with the cumulant and mutual correlation norms performing comparably at 0.2038 and 0.2052 m$E_h$, respectively, and mutual information marginally higher at 0.2473 m$E_h$. A consistent trend across all descriptors is an elevated error in the pre-equilibrium regime, where bond lengths fall below approximately 1.1 Å. This behavior is consistent with the feature analysis presented in Figure 2, wherein orbital entanglement values are substantially reduced at short bond lengths, providing the model with a less informative input signal in this region. Beyond the equilibrium geometry, errors decrease and remain well within sub-chemical accuracy across the full dissociation curve. Thus, we have selected the two-cumulant as the orbital correlation applied in the OrbGNN development due to its consistency and performance across the full dissociation of $N_2$. As shown in Figure 3, the two-cumulant provides a considerably tighter distribution of errors across bond lengths, rarely exceeding 1 kJ mol$^{-1}$.

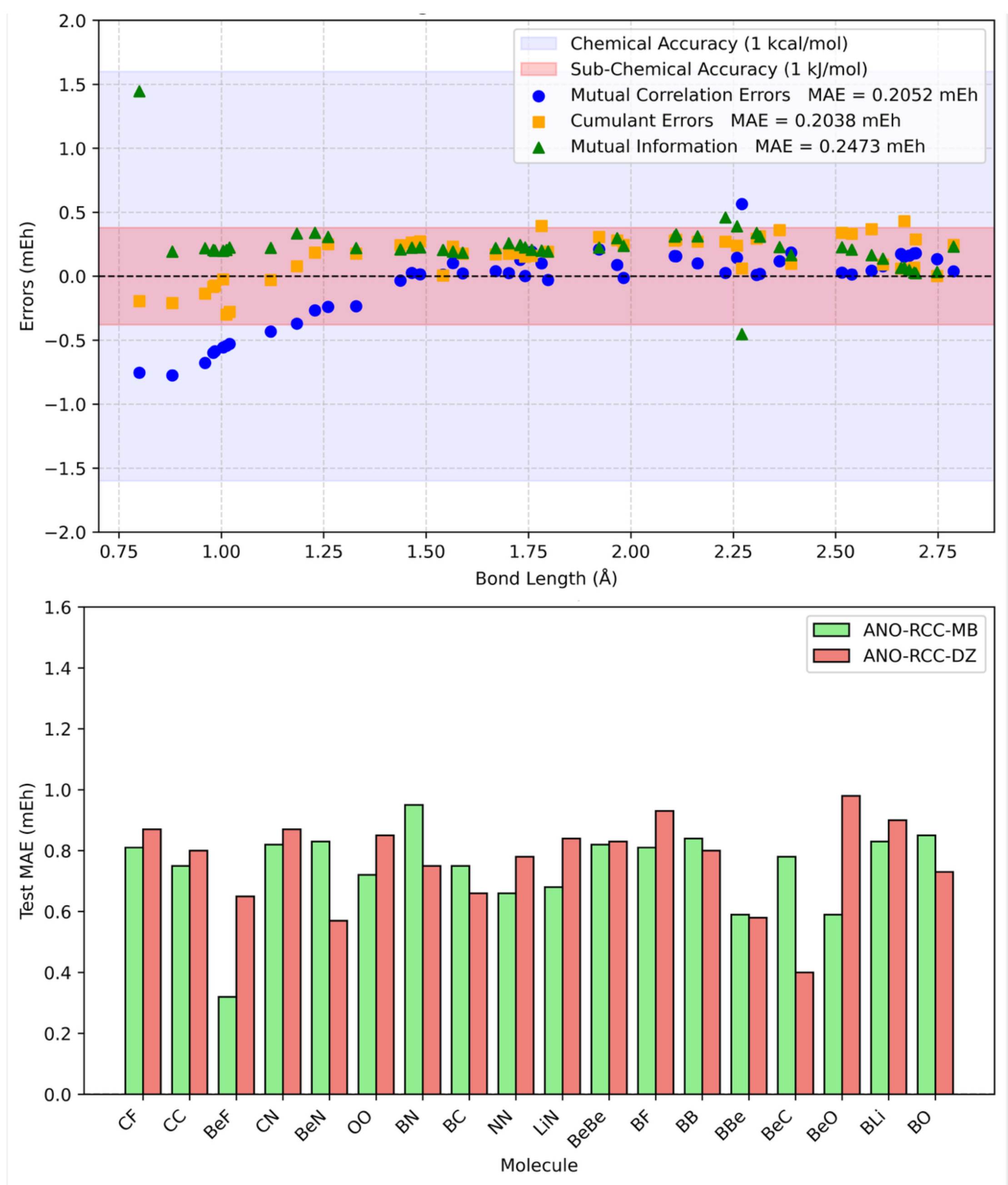

**Figure 3.** Top: OrbGNN accuracy as a function of bond length for the $N_2$ dissociation curve, comparing the three orbital connectivity descriptors used as edge features: the mutual correlation (blue circles), two-cumulant norm (orange squares), and mutual information (green triangles). The shaded region denotes the sub-chemical accuracy threshold of ±1 kJ mol$^{-1}$ (±0.38 m$E_h$). Bottom: Per-molecule test MAE across all 18 diatomic molecules for the ANO-RCC-MB and ANO-RCC-DZ basis sets, demonstrating consistent generalization of OrbGNN across molecular composition and basis set choice.

The lower panel of Figure 3 reports the per-molecule test MAEs for the ANO-RCC-MB and ANO-RCC-DZ basis sets. Across all 18 diatomic species (CF, CC, BeF, CN, BeN, $O_2$, BN, BC, $N_2$, LiN, $Be_2$, BF, $B_2$, BBe, BeC, BeO, BLi, BO), the test MAEs fall predominantly within the 0.6 - 0.9 m$E_h$ range for both basis sets, indicating consistent performance across chemically diverse bonding environments. The overall MAEs are 0.742 m$E_h$, and 0.773 m$E_h$ for ANO-RCC-MB and ANO-RCC-DZ, respectively. No individual molecule shows a pronounced deterioration in accuracy, suggesting that the orbital graph representation captures transferable electronic structure features across the different bonding motifs represented in the dataset. The results obtained with the minimal and double-zeta basis sets are broadly comparable, although ANO-RCC-DZ yields slightly larger errors for several species, likely reflecting the additional orbital flexibility associated with the larger basis. Overall, these results further support the use of the two-cumulant norm as the preferred orbital correlation descriptor for subsequent larger-scale training and demonstrate that OrbGNN maintains consistent accuracy across both molecular composition and basis set choice within this diatomic benchmark.

**OrbGNN: Octahedral Iron Complexes**

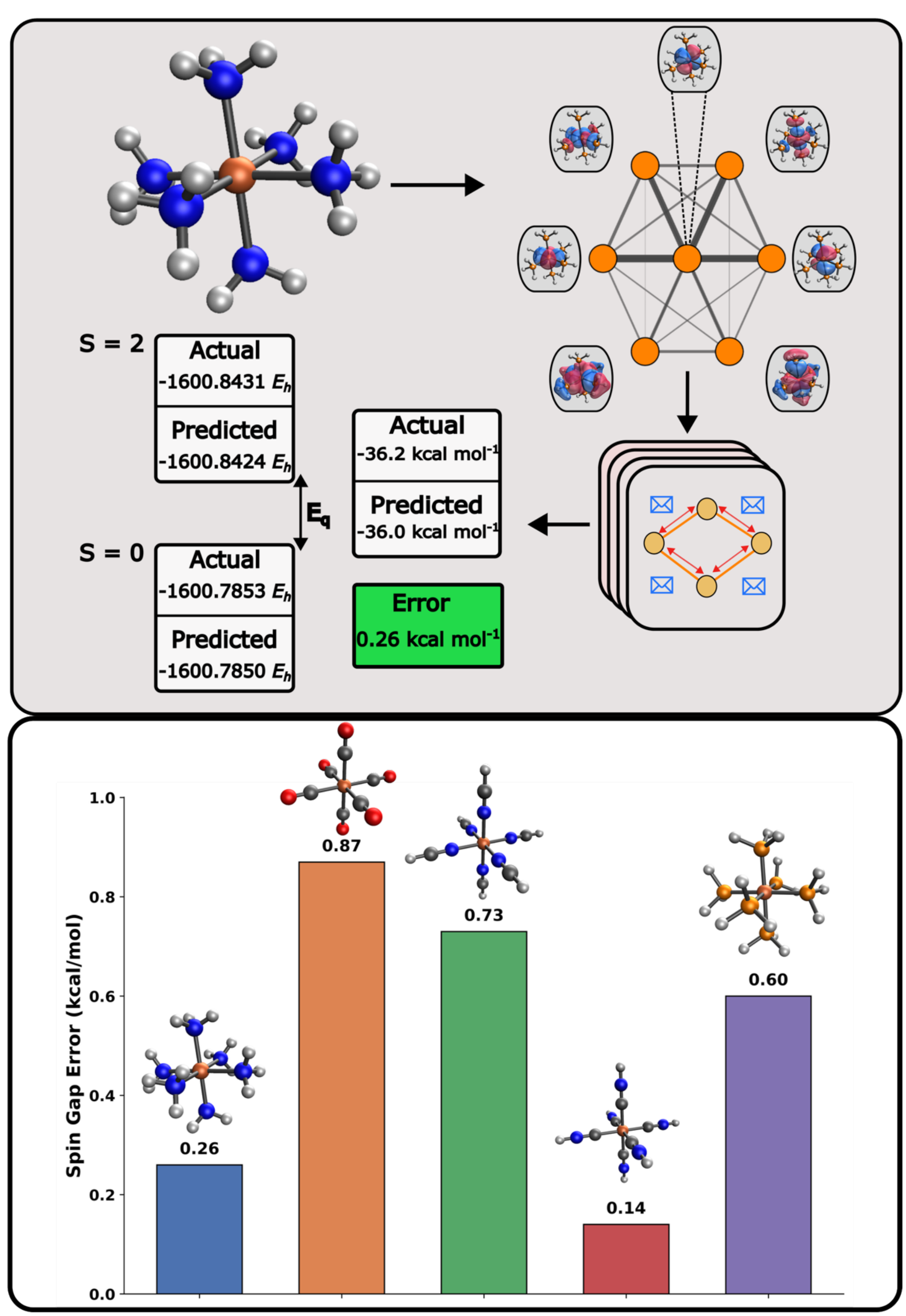

**Figure 4.** Fe(II) spin state energy gaps predicted by OrbGNN. Top: predicted spin state energy gap for $[Fe(NH_3)_6]^{2+}$. Bottom: summary of the predicted spin energy gaps for all five Fe(II) complexes.

To demonstrate the capacity of orbital graphs to accurately represent the electronic structure of transition metal systems, OrbGNN was rigorously evaluated on the prediction of correlation energies and spin gaps across a series of core Fe(II) octahedral complexes using benchmarks from Vlaisavljevich et al.[60] A robust training dataset was constructed by symmetrically stretching the six ligands along the Fe–L coordinate to map detailed potential energy surfaces, sampling four distinct active space sizes per complex to ensure the network could handle varying graph sizes. Five different $[Fe(II)L_6]^{2+}$ complexes were considered, where L = $NH_3$, CO, HCN, CNH, and $PH_3$. The molecules used for evaluation of the model's performance were taken at the equilibrium geometry for each system with a CAS(10,12) active space. The resulting model demonstrated exceptional fidelity to high-level quantum chemical references; as illustrated in Figure 4, OrbGNN produced spin gaps in good agreement with the CASPT2 reference values, achieving an MAE below 1 kcal $mol^{-1}$ across all five complexes. Using CASPT2 energies as a target for OrbGNN allows the model to obtain predictions for dynamic and static electron correlation. Although OrbGNN uses descriptors derived only from the active space, training against CASPT2 targets enables the model to learn energetic contributions that include both static and dynamic correlation, including contributions associated with orbitals outside the explicit feature space.

## Conclusions

In this work, we introduced OrbGNN, a graph-based architecture designed to accurately predict the energies of systems exhibiting static electron correlation. OrbGNN represents molecular orbitals as graph elements and captures their connectivity using electronic-structure descriptors, thereby incorporating physically meaningful information directly into the ML framework. OrbGNN derives its strength from directly encoding electronic structure information and, specifically, DMRG-derived orbital entanglement into the graph representation. This provides the model with a physically grounded description of multireference character rather than requiring it to infer such effects from molecular geometry alone. Combined with a graph-transformer message-passing architecture, this representation enables OrbGNN to capture orbital interactions across the electronic structure, providing a transferable framework for learning strongly correlated molecular systems. The results presented in this work demonstrate that orbital graph representations carry sufficient information to predict correlation energies across a range of molecular systems, from simple diatomic dissociation curves to complex open-shell transition metal systems, without requiring explicit knowledge of atomic coordinates or molecular geometry. On the diatomic benchmark, OrbGNN consistently achieves absolute errors within chemical accuracy across 18 distinct species and two different basis sets. Notably, the model successfully recovers accurate correlation energies in the strongly correlated stretched bond regime, which is a regime in which traditional single-reference methods qualitatively fail. This success establishes that the two-cumulant norm provides a physically meaningful encoding of orbital correlation, capturing the essential multiconfigurational character of the wavefunction in a form that is readily learnable by a message-passing architecture. This capability scales effectively to the substantially more demanding Fe(II) octahedral complexes, which present a rigorous test due to their large active spaces and intricate electronic structures. For the $[Fe(L)_6]^{2+}$ complexes, OrbGNN successfully navigates these challenges by distinguishing high-spin and low-spin configurations from orbital-level features alone, yielding precise correlation energies and correctly predicting spin-state ordering.

Building upon these foundations, future efforts will first focus on expanding the scale and diversity of the training data beyond localized potential energy surface scans. Broader coverage of chemical space, spanning diverse metal centers, oxidation states, ligand environments, and active space compositions, is necessary to maximize transferability, and the systematic nature of orbital graph generation makes this large-scale data collection highly tractable within existing high-performance computing workflows. To work towards this target, the implementation of automated active space searches with larger transition metal databases is a direct step forward. Additionally, future work will extend OrbGNN from a single-point property predictor toward molecular simulations by exploiting the close relationship between learned electronic structure information and existing equivariant MLIPs. This full integration allows for simultaneous learning of atomic and electronic structure, opening the door to accurate and comprehensive predictions of electronic properties and atomic potentials, namely geometry optimization and electronic correlation. Finally, this project opens the door to a subsequent study in which additional orbital information is provided for both the valence and virtual orbital spaces. CASPT2 target predictions could, in theory, be further advanced through the addition of orbital information outside of the active space.

## Acknowledgments

The authors gratefully acknowledge the CAREER and CAS-Climate programs of the National Science Foundation for financial support of this work (Grant no. CHE-2143354). The authors acknowledge the Infrastructure for Scientific Applications and Advanced Computing (ISAAC) of the University of Tennessee for computational resources.

## Data Availability

All software components used in this study are open source and freely available. The PYTHON package used to implement and train our ML models is PyTorch,[61] available on GitHub.[62] The complete set of data and workflows required to reproduce all figures is provided in the same GitHub repository.